# VERIFICATION OF CONCURRENT ENGINEERING SOFTWARE USING CSM MODELS[1]

**Jerzy Mieścickl, Mikołaj Baszun**
**and Wiktor B. Daszczuk**
Warsaw University of Technology
Department of Electronics and Information Technology
Warsaw, Poland
**Bogdan D. Czejdo**
Department of Mathematics and Computer Science
Loyola University
New Orleans, LA 70118

## ABSTRACT

An engineering design process may involve software modules that can executed concurrently. Concurrent modules can be very easily subject to some synchronization errors. This paper discusses verification process for such engineering software. We present a method for verification that requires several steps. First, the state diagram models are constructed that describe the design iterations and interactions with the designer. Next, the state diagram models are transformed into concurrent state machines (CSM). After that, the CSM models are analyzed in order to verify their correctness. In this phase, the modifications are performed in necessary. In the last phase the code is generated. The tools to support our method can be called new concurrent CASE tools. Using these tools the engineering software can be created that is verified for correctness in respect to concurrent execution.

## 1. INTRODUCTION

An engineering design may involve many different methods and tools (Ertas, 1993). The design process very often is based on an iterative search method. The iterative process of engineering design can be significantly enhanced by an appropriate system.

The requirements for a visual interactive software system for engineering design (VSED) are given in (Baszun and Czejdo, 1995). Such a system should allow the designer to specify or change design decisions in any phase of the design process. Additionally, the need for software for engineering design is documented in (Baszun et al., 1996). However, interactions with the designer and concurrently executed modules can very easily contain some synchronization errors. Therefore this problem should be addressed while developing concurrent engineering software.

[1] This work was supported by Grant No. 8T11IC00708 for Polish State for Scientific Research.

In this paper we present a method for verification of concurrent engineering software. The method requires several steps as shown in Figure 1. First, the state diagram models are constructed that describe the design iterations and interactions with the designer. Next, the state diagram models are transformed into Concurrent State Machines (CSM) models (Mieścicki, 1992, a & b). After that, the CSM models are analyzed in order to verify their correctness. In this phase, the modifications are performed if necessary. In the last phase code is generated. The tools to support our method can be called concurrent CASE tools. Using these tools the engineering software is crated that is verified for correctness in respect to concurrent execution.

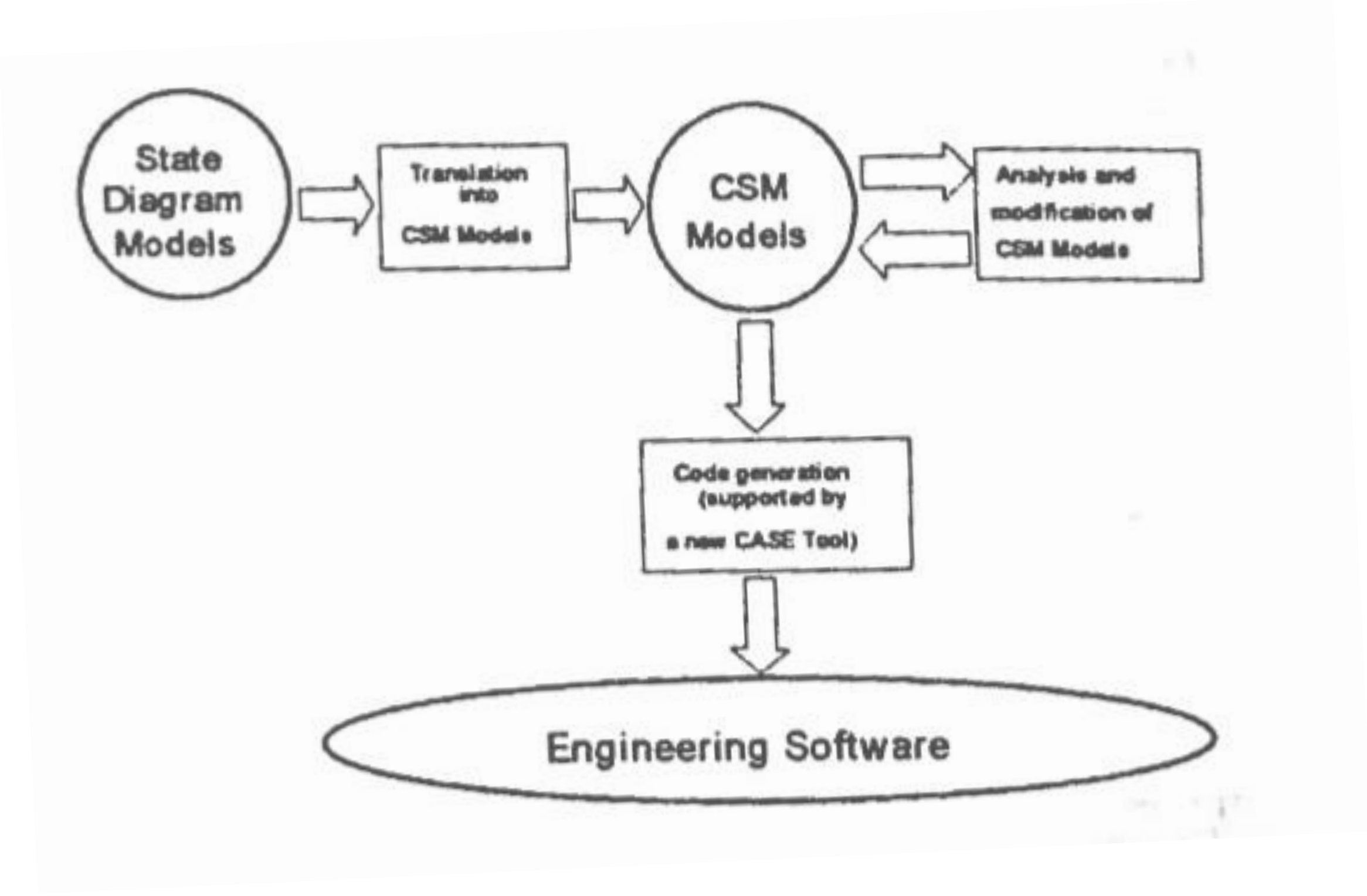

**Fig. 1. Software design methodology**

The paper is organized as follows. First, the characteristics of concurrent engineering software is described in Section 2. The transformation of state diagram models into CSM models is given in Section 3. Analysis and modifications of CSM models are discussed in Section 4.

## 2. CONCURRENT ENGINEERING SOFTWARE

### 2.1 MODELING AN ENGINEERING DESIGN PROCESS

An engineering system design process usually results in finding values of parameters fully describing the system. Nature of these parameters, called design parameters (DP) can be quite different: material, geometrical, electrical, architectural, etc. The design process should be preceded by the detailed analysis of physical phenomena outlighting detailed properties of the engineering system. Such analysis should lead to the creation of a physical model of the engineering system. The physical model includes the constraints on the design parameters (C_DP).

The application/market/utility requirements for the engineering system product result in some constraints on chosen properties of the designed system. In this paper, for simplicity, we assume that only one such property of the designed system is considered. We will refer to such property as output characteristics (OC). Constraints on output characteristics will be called C_OC. The criterion of evaluation of the quality of the output characteristics is based on comparison with the given ideal output characteristics (I_OC). To measure the closeness of these two characteristics an objective function (F_OF) will be used.

The designer must be given or must develop a mathematical model of the engineering system, giving complete mathematical relationships between all interesting quantities describing its internal and external behavior. The complete mathematical model also includes the precise algorithms of resolving of all mathematical formulas and equations describing the required relationships. Such mathematical model (F_OC) enables computation of the output characteristics for any set of values of design parameters.

It is a choice of the designer to use the proper design methods and accuracy. The choice affects the effectiveness of the design. Especially important may be planning of detailed strategy of searching for the final values of design parameters during design process. In many situations the best methods are based on an iterative process to find the final values of design parameters.

The design process could be treated as searching for final values of design parameters. Each parameter may have numerical value belonging to some set of allowable values. Very often such set includes all numbers or real numbers restricted to a particular range. It means that the values of parameters may need to be searched from the infinite scope of possible values. One feasible approach is to use some kind of iterative search. In an iterative search a strategy of searching for the final design (A_DP) needs to be specified. A simple example of such strategy is a sequential search algorithm with the specification for changes of design parameters DP. The initial value of DP is specified first by the designer. In each cycle of the search the output characteristics and objective functions are computed. The process is terminated when the minimum value of the objective function is found. The search strategy should guarantee the global minimum of an objective function (OF) rather than its local minimum.

Because of repetitive nature of data access and updates during iterative design process, it is necessary to store the complete information about the current design stage (current iteration). We will introduce new symbols to describe more explicitly the current stage of the design process:

- CC DP - logical data resulting from DP constraints checking. Such checking can be for simplicity treated as the comparison of the relevant data from data blocks DP and C_DP. This operation needs to be done after every change of DP.
- CC OC - logical data resulting from OC constraints checking.
- M_OF - logical data which indicate when the objective function has obtained the minimum value.

Here we assume that all numerical and logical data store both the current and historical values. This is an important feature of an design process because the designer has always access to information about any process stage and he/she can restart the design process from any stage using selected values of parameters and functions.

### 2.2 STATE DIAGRAM MODELS FOR ENGINEERING SOFTWARE.

The engineering software to support an engineering design can consists of several modules. In particular, the software for the interactive design described in Section 2.1 can consist of many modules. Here for the clarity of presentation we will focus our attention on interactions of only two modules: a Designer's Interface Module, and a Computations' Control Module, what is shown in Figure 2.

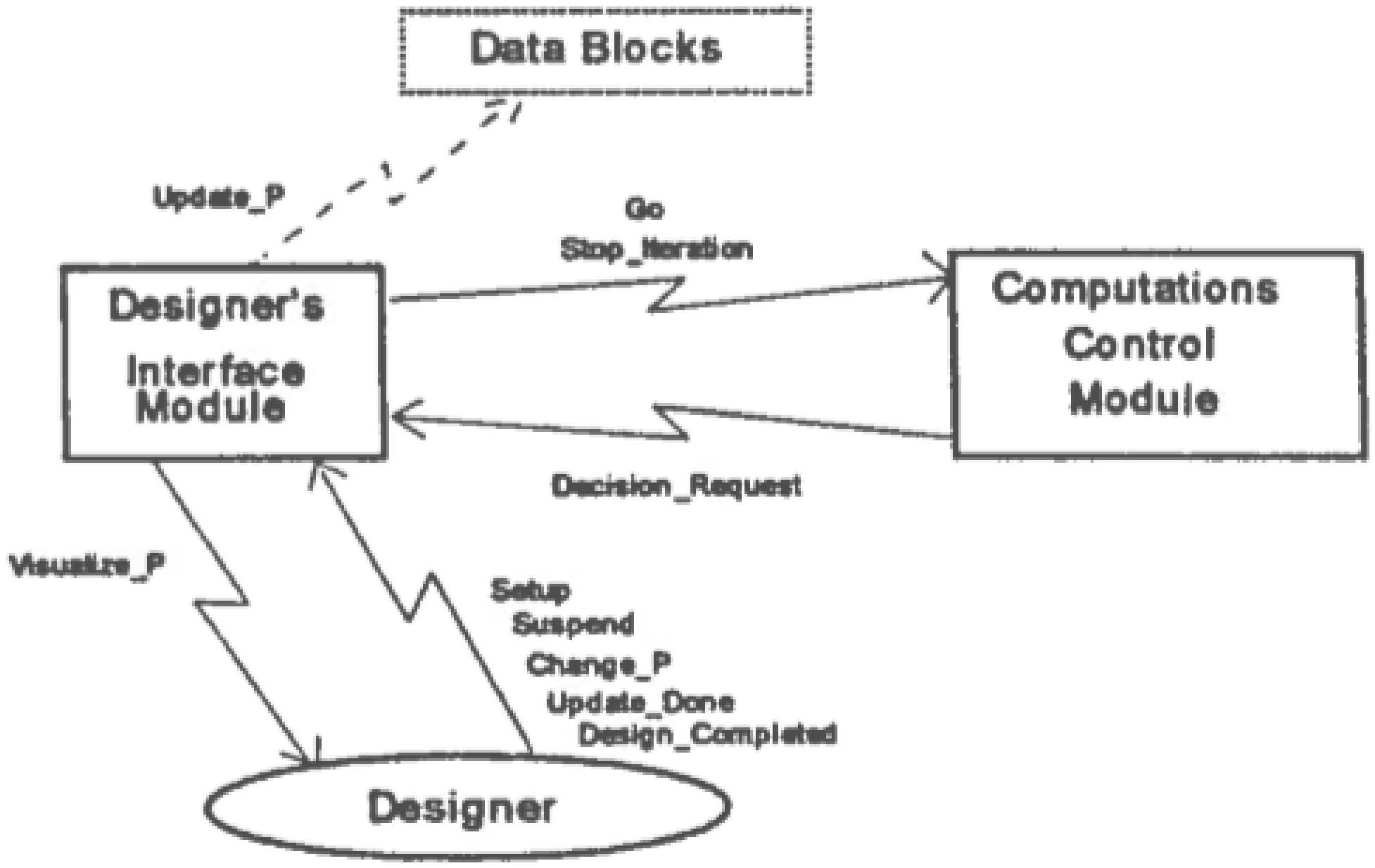

**Fig. 2. Module Interactions design**

The Designer’s Interface Module allows the designer to provide initial values for design parameters and constraints on design parameters (*Setup* request). The initial formulas for output characteristics and constraints on output characteristics, the formulas for computation, and method of search and search increments of objective functions are provided by the designer also via Visualization Module (*Setup* request). Request *Change_P* is to change these parameters and *Update_Do*ne request informs the system that it can proceed with automatic iterations. The Suspend request stops the automatic and allows the designer to change the parameters again using *Change_P*. The visualization of the design is available for the designer. The request *Design_Completed* is issued when the designer is satisfied with the design. The Internal Computations Module receives requests *Go* and *Stop_Iterations* from the Visualization Module. It can also send the request *Decision_Request* when automatic computations cannot proceed.

During the interactive design process the Designer’s Interface Module displays the complete information about the design process (*Visualize_P*). If necessary, the designer can modify initial values for design parameters and constraints on design parameters using that Module. The formulas for output characteristics, constraints on output characteristics, the formulas for computation of objective functions, and the method can be modified using also this Module. In such system the designer can direct and assist in any phase of the continuous design process.

Description of all software modules will be done using a behavioral model (Embley et al., 1992; Rumbaugh et al.,1991). This model is an extended state diagram that describes the dynamic behavior of objects. It has three basic components: states, for each object in the given class, triggers that cause the transition of an object from one state to another and actions performed during the transition. Triggers can be either Boolean conditions or events.

### 2.3 DESIGNER'S INTERFACE MODULE

Figure 3 shows a simple object-behavior model for the Designer's Interface Module. Each state is indicated by a rounded box and each transition is indicated by an arrow with a label. The first part of label (before slash) specifies the trigger and the second part (after slash) specifies the action to be performed during transition. Boolean conditions (if present - there are no Boolean conditions in Figure 3) are indicated by brackets while for events the brackets are not used.

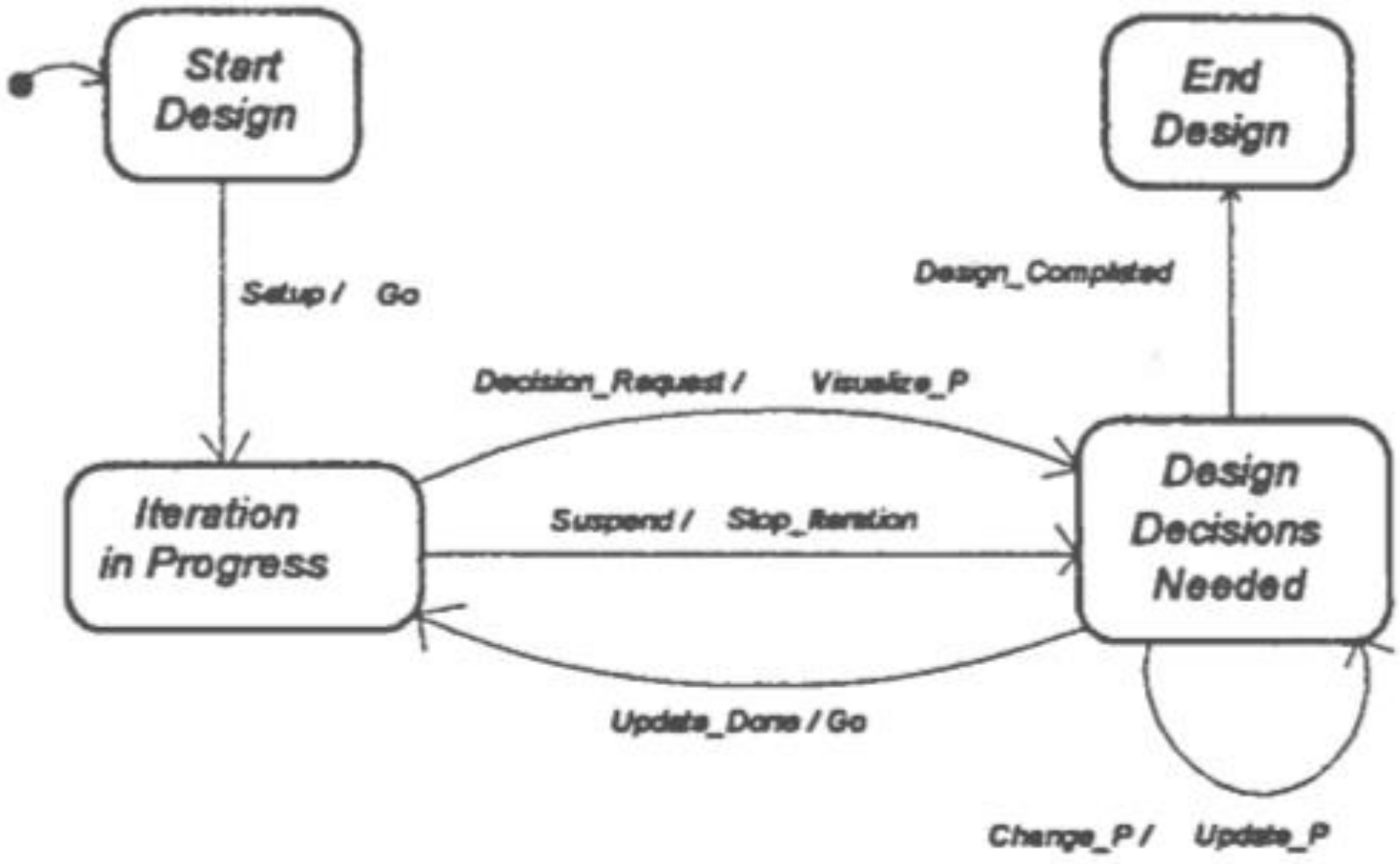

**Fig. 3. State diagram of Designer's Interface Module**

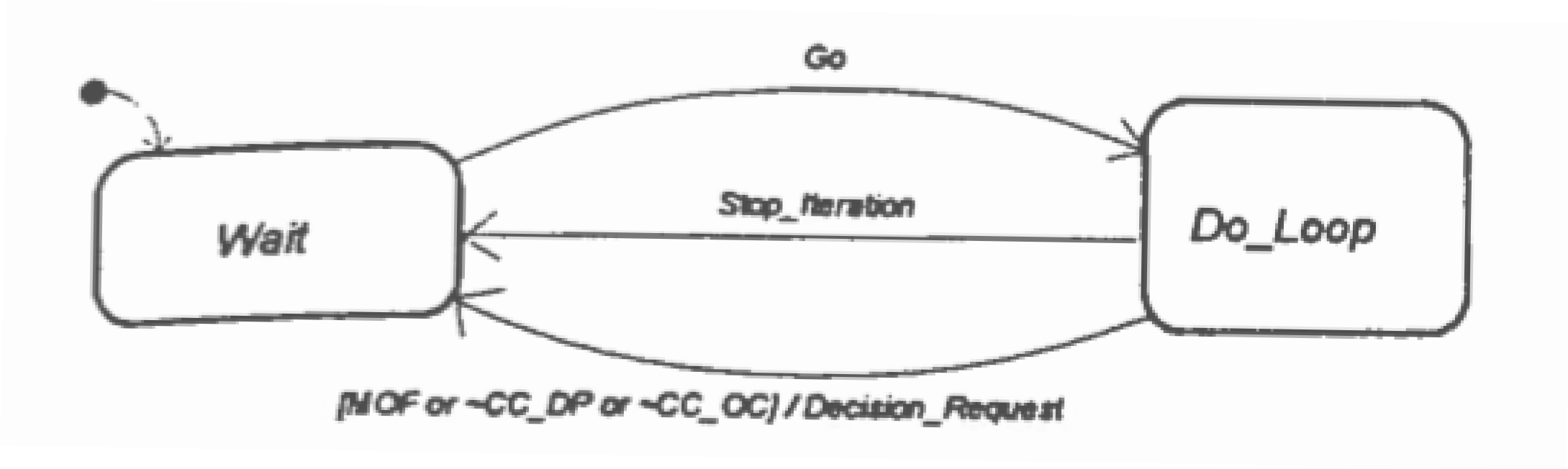

**Fig. 4. State diagram of Computations Control Module**

Initially, Designer's Interface Module (while in **Start Design** state) receives the request (the *Setup* event) from the designer to start the design. That results in starting the design process (the event *Go* is generated and sent to Computations Control Module). The design process can be in two basic states corresponding to a passive role (the **Iteration in Progress** state) and an active role (the **Design Decision Needed** state) of the designer. It is assumed that in the first state all iterations are done automatically and designer an only observe the current results. The designer can issue a request (*Suspend* event) in order to stop automatic iterations and participate actively in the next phase of the design. When this request is received all automatic are stopped (event *Stop_Iteration* is generated and sent to Internal Control Module) and all visualization screens allow the designer to change appropriate design components. Another possibility for transition to the state corresponding to the active role of the designer is related with the situation in which a minimum of objective function is found or automatic iterations run into problem (the event *Decision_Request* was received from Internal Control Module).

The second state corresponds to the active role of the designer (**Design Decision Needed** state). The designer can change all design parameters, constraints etc. using *Change_P* request. Once all modifications are done the designer can issue a request (*Update_Done* event) in order to start automatic iterations. More specifically, when this request is received the request to start automatic iterations is issued (event *Go* is generated and sent to Internal Control Module), and the state is changed to one corresponding to the designer passive role. If the designer, while in active mode, is satisfied with the design he can issue a request to end the design (the *Design_Completed* event causing transfer to the **End Design** state).

## 2.4 COMPUTATIONS CONTROL MODULE

Computations Control Module specifies the automatic design iterations. Initially, (while in **Wait** state), it receives a request to start the iterations (*Go* event). That causes a transition to a state (**Do_Loop** state) that enables a sequence of operations defining iterations. If, during the execution of this sequence of operations the constraints on design parameters are not satisfied (the condition CC DP is false) or constraints on output characteristic are not satisfied (the condition CC_OC is false) or the minimum of objective function was obtained (the condition M_OF is true) then the iterations stop (the transition take place to the Wait state ) and the appropriate message is sent to the **Wait** state) and the appropriate message is sent to the Designer 's Interface Module (the event *Decision_Request*). The automatic iterations also stop if the request from the designer's Interface Module) was received to stop iterations (event *Stop_lterations*).

The **Do_Loop** state is an abstract state that can be defined by another diagram. It should specify the automatic design iterations that include a following sequence of operations:

1/ checking the constraints on design parameters (operation UpdateCC_DP(DP, C_DP); )
2/ computing output characteristic (operation UpdateOC(DP,F_OC) );
3/ checking the constraints on output characteristics (operation UpdateCC_OC(OC,C_OC) );
4/ computing objective function (operation UpdateOF(OC,I_OC,F_OF) );

# 3. TRANSFORMATION OF STATE DIAGRAMS INTO CSM MODELS

## 3.1 CONCURRENT STATE MACHINES AND THEIR BEHAVIORAL PATIERNS

Concurrent State Machines (CSM), introduced first in (Mieścicki,1992, a & b), resemble well known finite state machines (more specifically: Moore-type finite automata) with their finite sets of states, finite input and output alphabets, next-state relations and output functions. However, in a CSM symbols from its input alphabet an occur independently of each other. This way, for the machine M with the input alphabet e.g. A = {*a*, *b*} at any instance of time we can expect one of four possible input situations: either no symbol comes, or *a* alone, or *b* alone or both *a* and *b* concurrently. Each of these possibilities is simply a subset of A: in our example the empty subset ∅, {*a*}, {*b*}, {*a, b*} respectively. Consequently, transitions among states are defined as the results of the reception of specific subsets of A, while in conventional finite state machines transitions are performed in response to single symbols from A.

Similarly, machine's output function maps the set of states into the set of subsets of the output alphabet. In other words, in any state the machine can produce either no output symbol (i.e. empty set) or single symbol or any other subset of its output alphabet. Moreover, in contrast to finite state machines, it is not required that input and output alphabets are disjoint. Thus, while the conventional finite state machines can only passively wait for new input from the environment, CSM can produce its own output symbols that are immediately 'audible' to itself and may cause spontaneous transitions between states.

Concurrent State can be organized into systems of CSMs. Important assumption is that within the system two following rules define the communication among system components:

- all sets of symbols produced at a given instance of time by any source (i.e. by the system environment as well as any component of the environment) are composed into the 'global' system input which is set union of these individual sets,
- at any time the resulting global set of symbols is immediately and faultlessly broadcasted to all parts of the system.

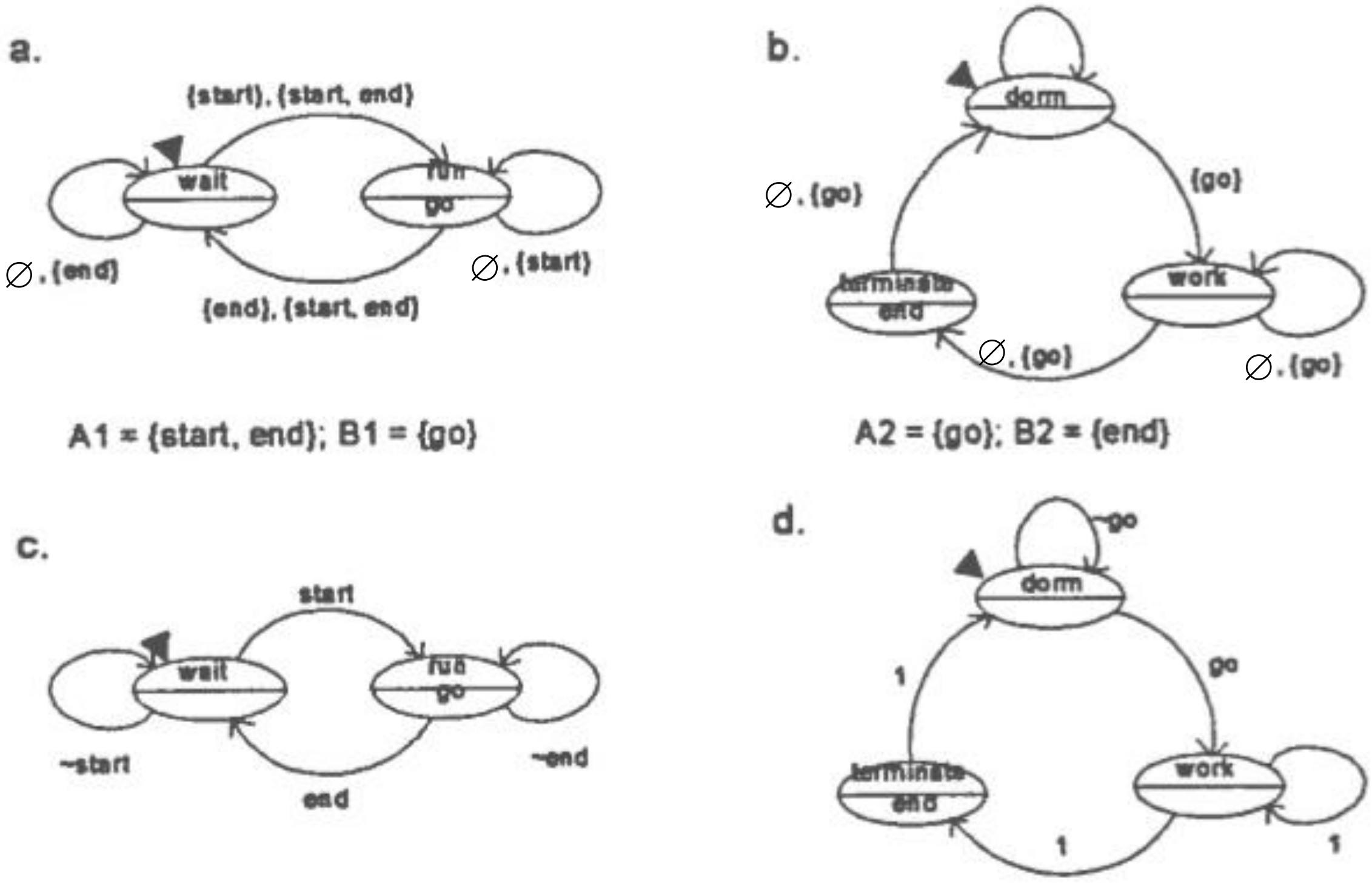

**Fig. 5. Example CSMs: M1 and M2 (a, b, respectively) and their behavioral patterns (c,d)**

More detailed discussion of formal properties of CSMs falls beyond the issues discussed in this paper. We confine ourselves to the informal example aimed to support the reader's intuition. In Figure 5a the simple two-state machine (M1) is shown. Its input alphabet is A1 = {*start*, *end*}, the output alphabet is B1 = {*go*}. The initial state is **wait** and in this state the machine produces an empty output. The machine has to remain in **wait** state until it receives the empty input ∅ or {*end*}. Receiving either {*start*} or {*start*, *end*} the machine has to pass to **run**. We may say that in **wait** the machine ignores the *end* symbol: indeed, its presence or absence does not influence the machine's behavior. However, *start* is carefully watched: if it does not come - the machine rests in **wait** state, if it comes (alone or coincidently with *end*) - the machine changes its state to **run**.

By the similar argument, in the **run** state the machine ignores *start* and is sensitive to *end*. Machine M1, as long as it is in **run** state, keeps producing its output symbol *go*.

The rules described above are conveniently represented by behavioral patterns of both machines, shown in Figure 5c and 5d, respectively. Note that in behavioral patterns, expressions labelling the edges are no longer sets of subsets of input alphabets but Boolean formulas of very clear and natural interpretation. Now, for **wait** state of M1 we see that if *start* does not come - the machine has to stay in **wait**, if *start* does come - it changes the state to **run** etc. Boolean formula **1** means practically 'always' (or: regardless of the input). Similarly, formula **0** would mean 'never'. Normally, it is not used in the graph because transitions labeled with **0** are never executed and should be simply removed from the graph.

It should be emphasized that in the general case the more complex Boolean formulas can be used, composed with the use of Boolean operators like the negation, logical sum and logical product.

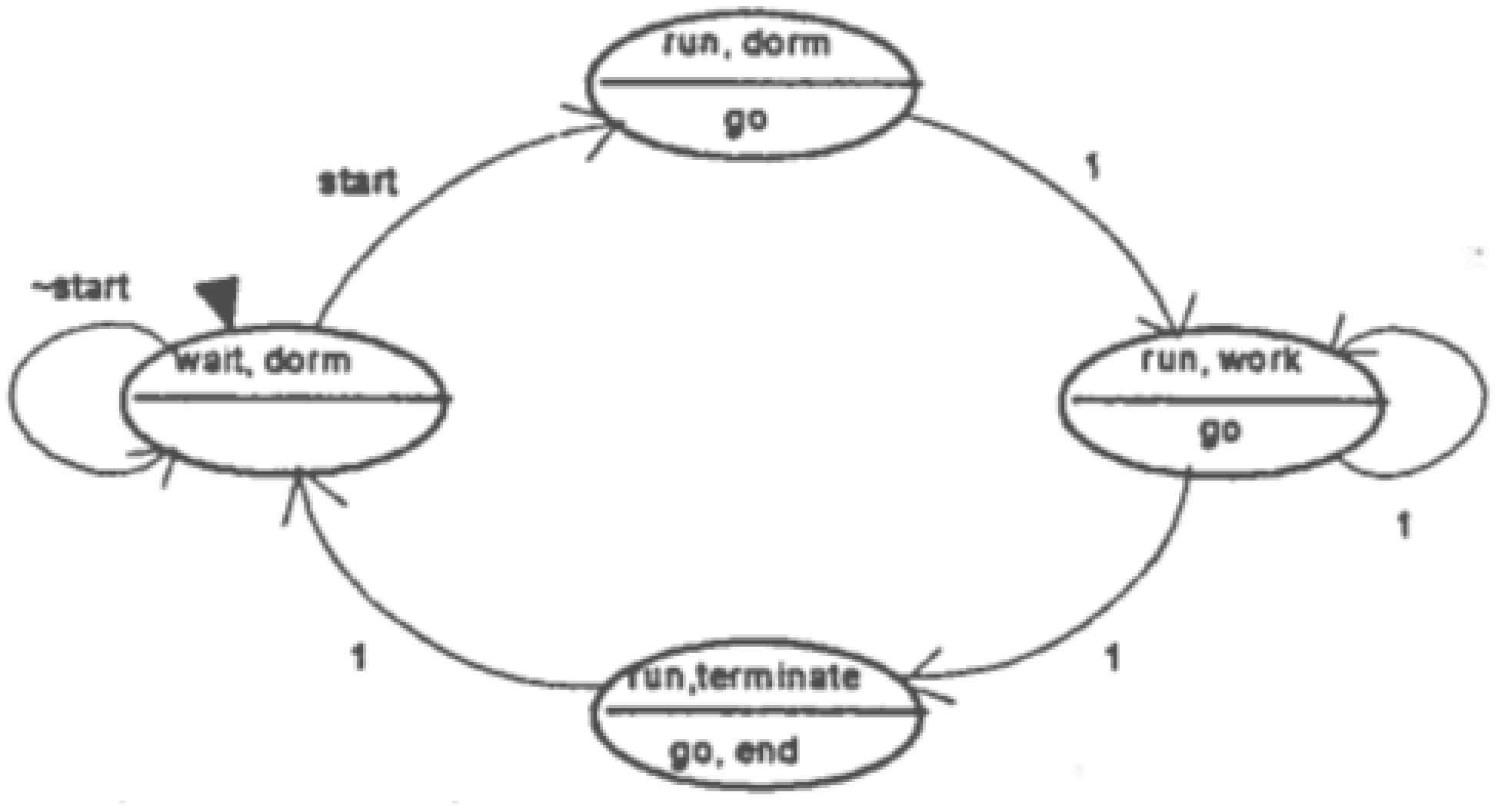

**Fig. 6. Behavioral pattern of a system of machines from Fig. 5**

Behavioral patterns of CSMs provide the convenient shorthand notation for machine's input behavior it is not the only advantage of this important notion. It can be shown (see Mieścicki, 1994) that the transitions between the system states (defined as the vectors of states of system components) are labelled with Boolean products of relevant formulas from the individual patterns. This provides the method for obtaining the graph of system that are reachable from the initial state.

To do this one has to compute Boolean products for all potentially possible transitions from system's initial state and to take into account only non-0 transitions. One of these transitions may lead back to the initial state, but normally also new reachable states are obtained. Then, the procedure should be repeated in consecutive steps for all new reachable states until no new reachable states result. The reader is encouraged to check that the procedure applied to the system consisting of M1 and M2 results in the graph shown in Figure 6.

It is important to note that in practice, when the analysis of system's behavior is the primary issue, detailed graphs of concurrent state machines (as in Figures 5a and 5b) are hardly necessary. Behavioral patterns of system components (as in Figures 5c,5d) and the resulting behavioral pattern of a system (as in Figure 6) provide the necessary information on the system behavior (Daszczuk, 1992; Kołodziejak, 1992).

The software tool named COSMA has been implemented in the Institute of Computer Science, Warsaw University of Technology[2], that supports defining and modifying the Concurrent State Machines as well as obtaining reachability graphs for systems of CSMs. In Chapter 4 we will show how to use this software

[2] COSMA has been implemented by Andrzej Lachowski as a part of his MSc thesis advised by Jerzy Mieścicki

for the purpose Of the analysis of the engineering software described in Chapter 2. The general idea outlined above can be significantly improved by introducing Boolean formulas for outputs from states and by defining algorithms for abstractions and refinements in CSMs. This is beyond the scope of this paper.

### 3.2 PEPRESENTIN STATE DIAGRAMS BY CSMs

The meaning of a state diagram can be generally explained using Figure 7:

- stay in **state1** (and execute activities relevant to it) until neither *event1* nor *event2* occurs, nor *condition3* is true,
- on *event1* evaluate *condition1* and if it is true - execute *action1* and then change to **state2**, otherwise remain in **state1**,
- on *event2* change to **state3** (producing no action during the transition),
- execute *action3* and pass to **state4** immediately if *condition3* becomes true.

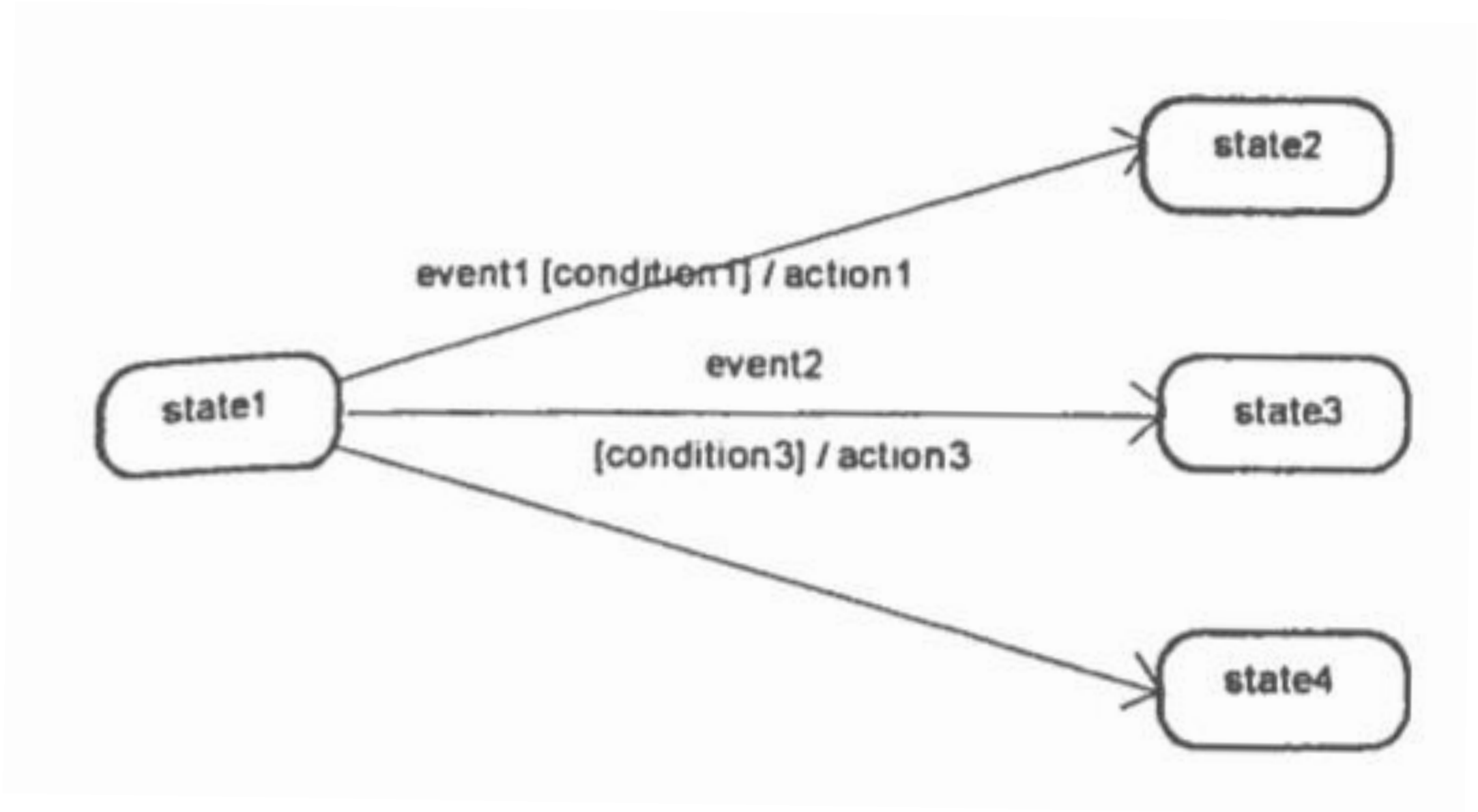

**Fig. 7. Fragment of state diagram**

The rules given above are easily converted into a behavioral pattern of a CSM. State diagram states remain CSM states, events and conditions become CSM input symbols while actions (generated upon the transition) become CSM output symbols. However, in CSMs output is produced not by but by states therefore in a CSM model some additional states are necessary, one per each edge in which action is generated.

Note that the state diagrams do not define clearly what would happen if two or more events occur in the same time. Usually it is assumed either that the events are 'very short' and their coincidence is unlikely, or that they are somehow ordered (e.g. by underlying system software or other-type 'synchronizing demon') so that we practically deal always with only one event occurrence in our application level. In contrast to this, CSM model supports the coincidences of input/output symbols as well as single event occurrences. Indeed, using the appropriate Boolean formulas one can 'openly' specify pattern's response to any set of events, for instance *ev1 and not ev2*, *not ev1 andev2* and *ev1 and ev2* can be three formulas representing three different transitions in the CSM graph and no underlying assumptions nor ‘synchronizing demons' are needed.

Additionally, we decided to enhance the protocol of communication among system components. We assume that any action (i.e. CSM output symbol) which is supposed to be an input event for some other system component is in fact the message between two system components and it be acknowledged by the recipient prior to moment when both communication partners proceed with their further activities. Thus, the reception of the input event (e.g. *event1*) has to be acknowledged by sending the appropriate ACK output (e.g. *ACKevent1*) while sending the output action (e.g. *action1*) should be followed by waiting for the appropriate acknowledgement (e.g. *ACKaction1*) from the supposed recipient of the message. However, for conditions as well as for events that come from the environment (not from other system components under consideration) sending acknowledgements is not required. Also, it is not necessary to wait for an acknowledgement after sending the output action which is not 'consumed' by other system component but goes out of the system beyond the modelled environment.

The communication protocol outlined above significantly increases the number of states in the CSM model as compared with the state diagram. Nevertheless, we have decided to apply it, partly in order to make the among models more realistic, partly because just the preliminary experiments have shown that without this simple hand-shaking protocol the system soon becomes simply a mess.

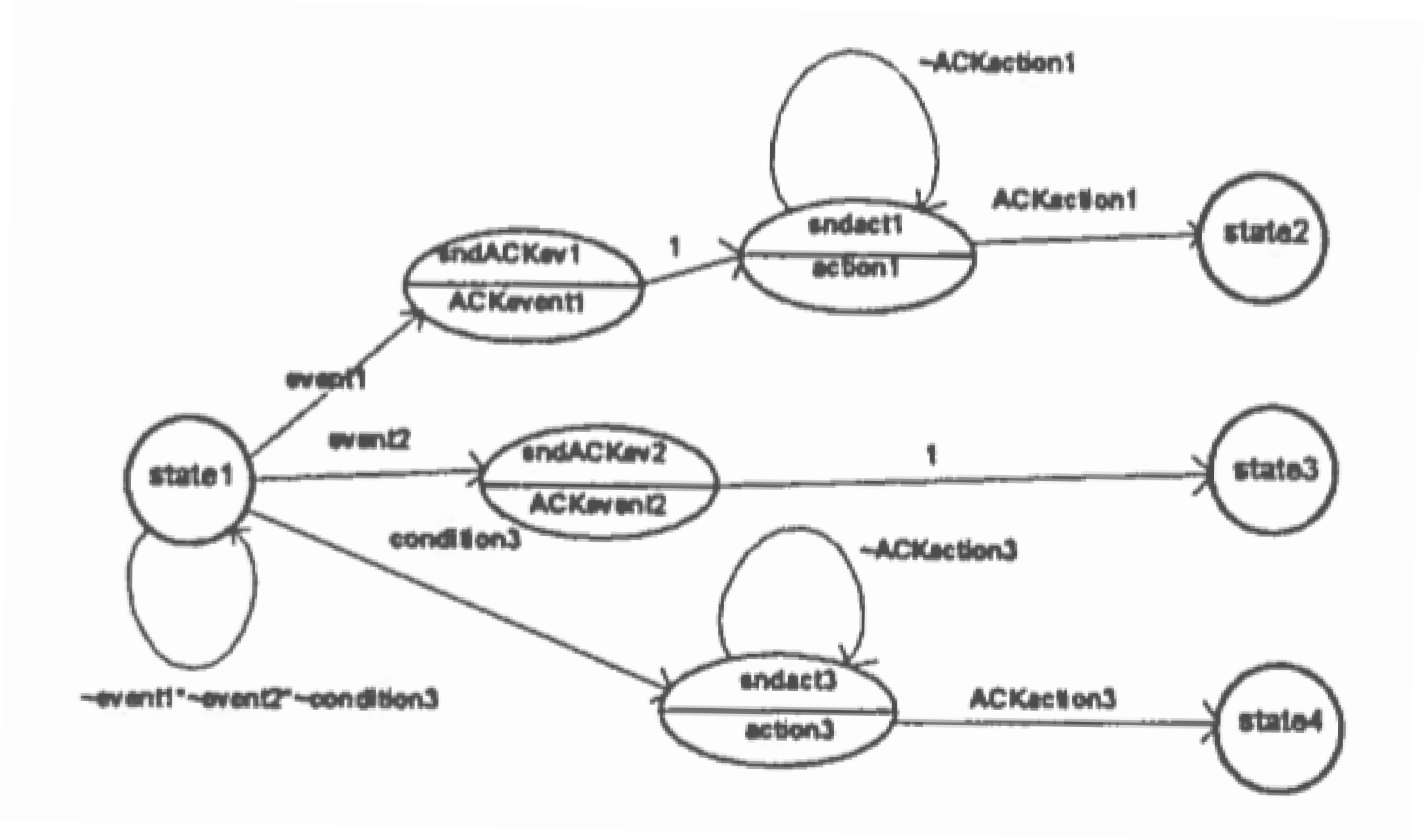

**Fig. 8. CSM model of the state diagram from Fig. 7**

Finally, the CSM model (more specifically, the CSM behavioral pattern) of the state diagram from Figure 7 is shown in Figure 8.

## 4. ANALYSIS OF THE CONCURRENT ENGINEERING SOFTWARE

By applying the above to example state diagrams of Designer's Interface Module (Figure 3) and Computations Control Module (Figure 4) we obtain CSM behavioral patterns as in Figures 9 and 10, respectively. The graphical conventions differ from the previous figures because of the properties of the present version of COSMA tool used for the specification of both models, however the principles incorporated in both models are quite easily seen.

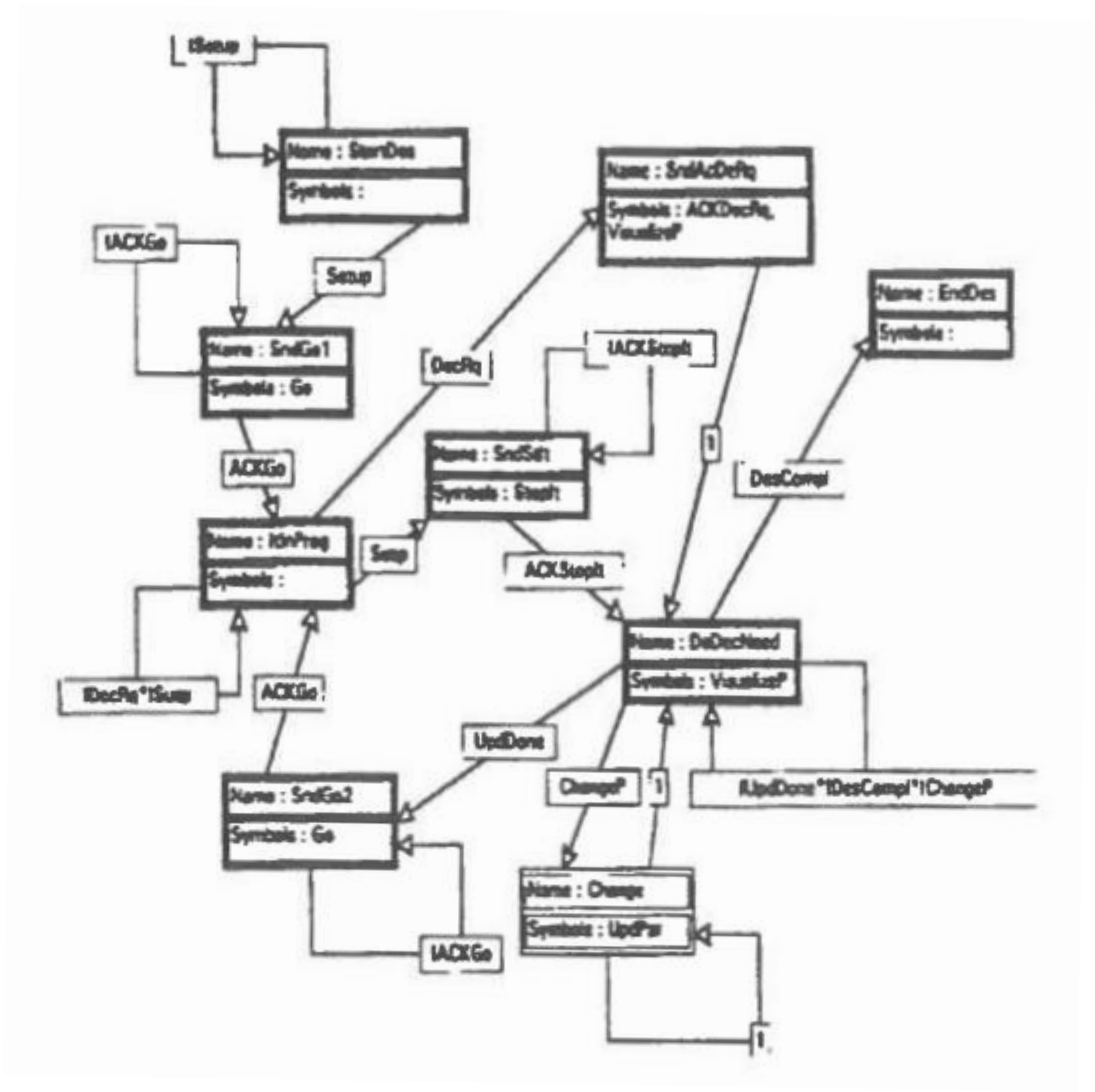

**Fig. 9. CSM model of Designer’s Interface Module**

States are rectangular boxes: upper part of a box contains state name while lower one contains output symbols produced by this state (if any). Remember that labels of transitions are Boolean formulas, where ‘!’ stands for negation, ‘*’ for Boolean product and '+' for Boolean sum. Formula '**1**' means 'unconditionally' or 'always'. Remember also that the CSM models allow the nondeterministic choice: if at some state two (or more) Boolean formulas are true - one of (non-deterministically selected) transitions is executed.

Note also that the hand-shaking protocol is applied to the messages exchanged between the two modules under concern (*Go -ACKGo, StopIt - ACKStopIt, DecRq - ACKDecRq*) while other symbols (from and to Designer and Data Blocks, see Figure 2) do not require acknowledgments.

Now, let the two CSM patterns shown be a concurrent system of two components, operating concurrently and asynchronously of each other. The Designer and Data Blocks (see Figure 2) make the system's environment. The graph of system's reachable States (obtained using the COSMA software) is shown in Figure 11.

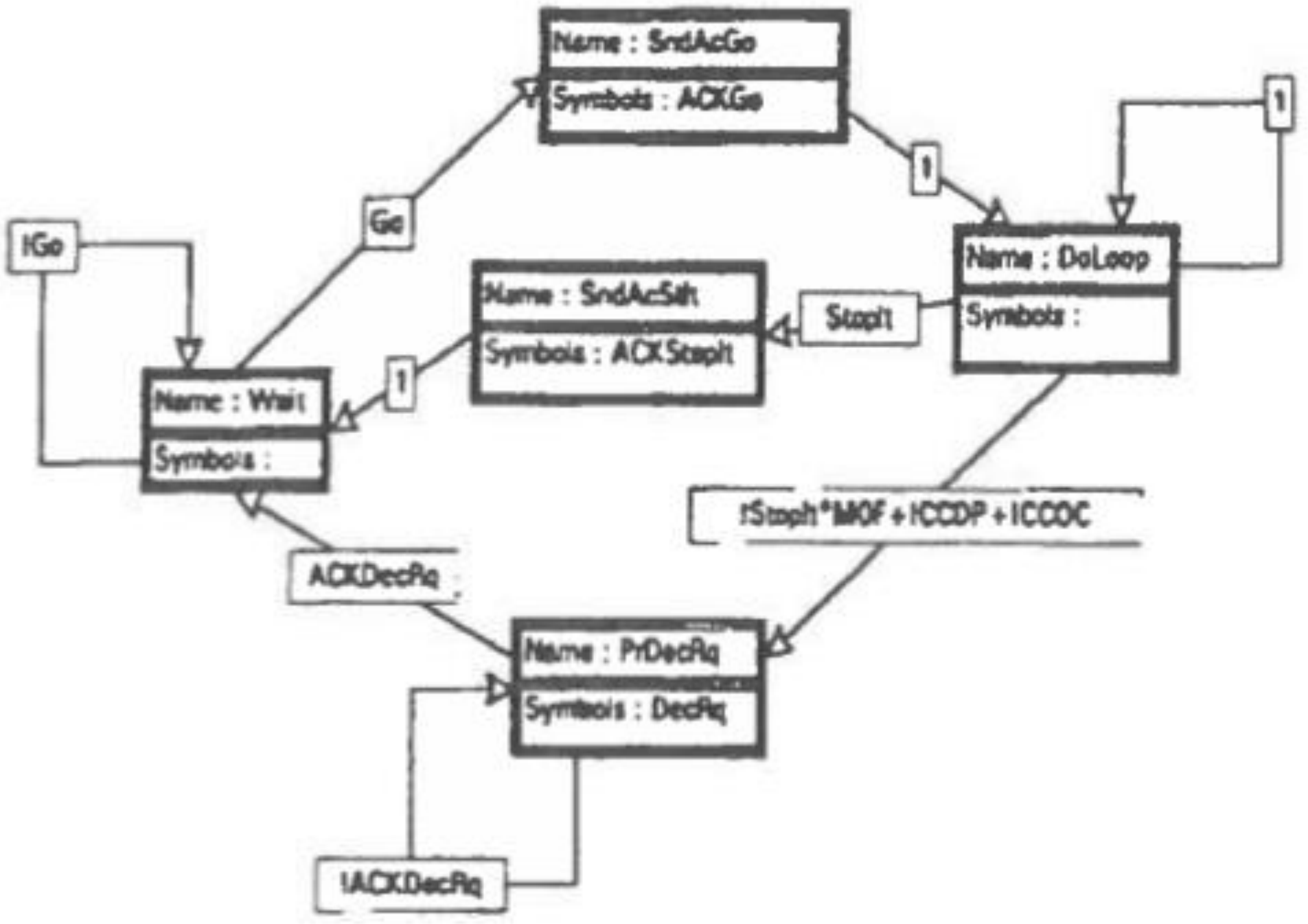

**Fig. 10. CSM model of Computations Control Module**

We can note first that while system components have nine and five states (respectively) - the system has as few as fourteen reachable states out of forty five that could be naively expected. Moreover, the inspection of the graph quickly reveals two 'suspected' system states which, once entered, cannot be leaved. One of them (**EndDes_Wait**, at the bottom of the graph) represents a harmless, predicted situation: it is entered upon receiving the *DesignCompleted* message from the Designer and means the end of whole design process. However, the other 'suspected' state (**SendStopIteration_ProduceDecisionRequest**, marked in Figure 11 with black markers), represents an actual system deadlock, caused by an unintended synchronization fault.

The analysis of three edges leading to this deadlock state reveal that the system components actually allow for the deadlock. It occurs when the for the designer's decision from Computations Control Module (due to logical condition computed within the iteration loop) coincides with the decision to suspend the computation that comes from the unaware designer. If such coincidence occurs - both modules wait for acknowledgments that are unlikely to come, thus the system becomes deadlocked.

Such a situation, once recognized, can be fixed in many ways. One can try, for instance, to modify the system components so that the deadlock state is never entered or so that the 'suspected' state, once entered, has to be immediately and unconditionally left. We made use of the latter solution. We have modified both CSM models so that the Designer's Control Module in **SendStopIteration** state produces not only *StopIteration* message but also ('for any case') *ACKDecisionRequest*, while the Computations Control Module in its **ProduceDecisionRequest** state issues ('for any case') *ACKStopIteration* in addition to normal *DecisionRequest* message. The resulting system graph, shown in Figure 12, is actually deadlock free. The reader is encouraged to check in this graph how the system components now interact with each other and with the designer.

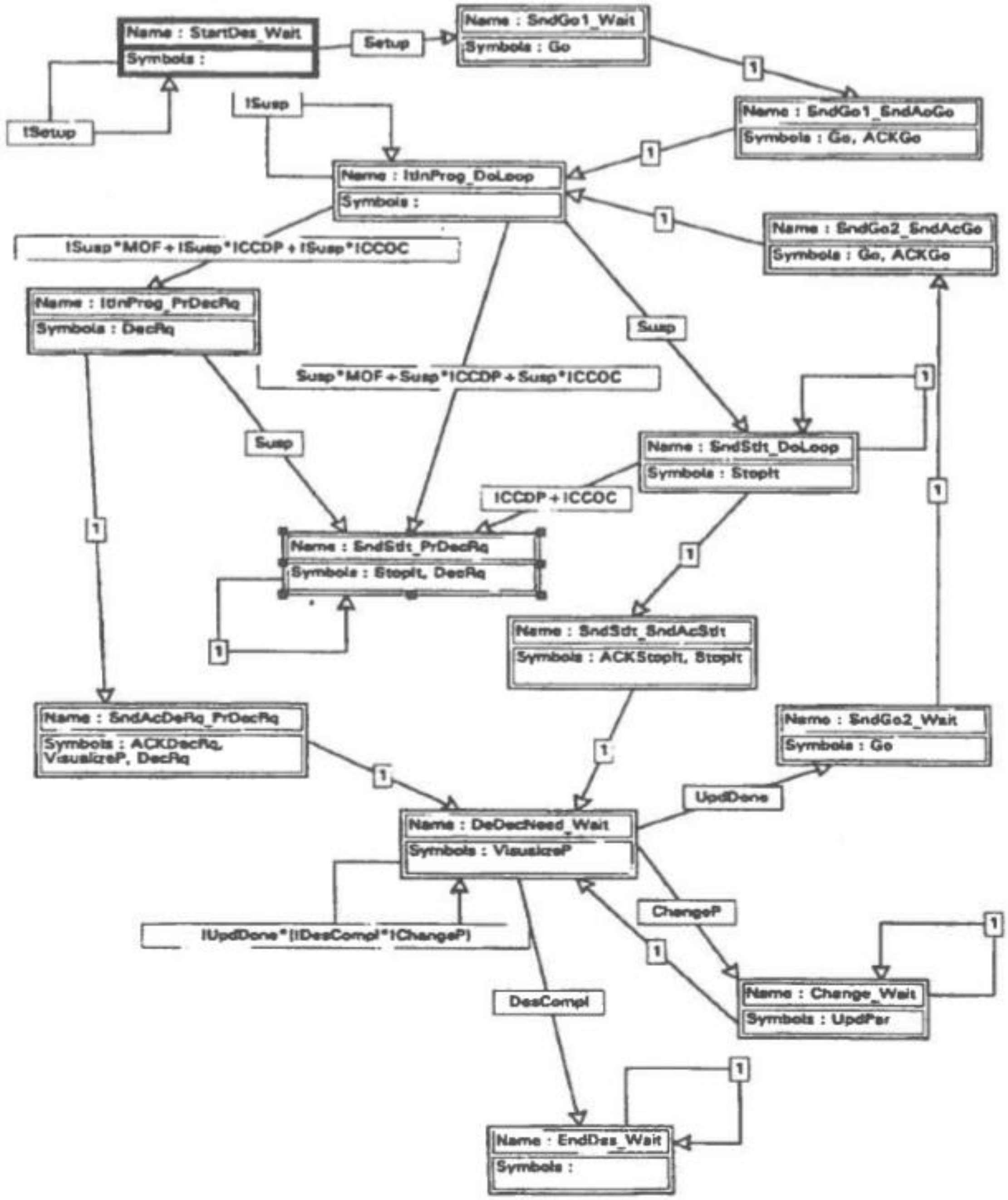

**Fig. 11. Graph of system's reachable states with a deadlock**

## 5. SUMMARY

In this paper we showed that the use of CSM models (and the software tool for obtaining their reachability graphs) can be effectively used for the improved specification of state diagrams behavior (e.g., by 'open' specification of the reaction for the coincident events or by introducing enhanced communication rules) as well for the analysis of the synchronization among the elements of a concurrent engineering software. By the inspection of the system's reachability graph we have encountered the unintended deadlock, quite obvious only after its discovery but hardly predictable in state diagrams or CSM models of individual system components. By analyzing paths that can lead to the deadlock we were able to find the remedy for it, to improve the system components and to prove that the system modified this way actually recovers from the deadlock.

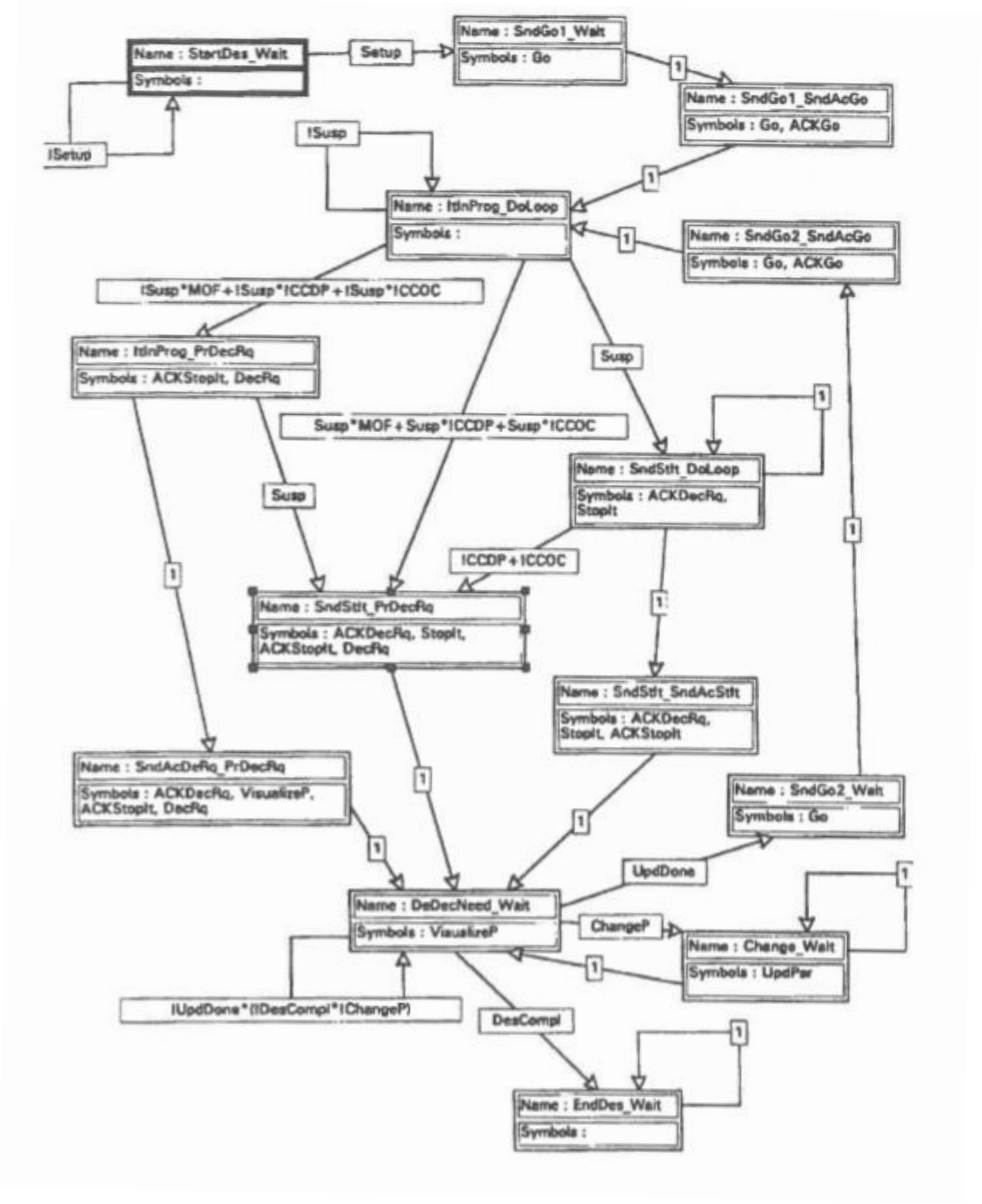

**Fig. 12. Graph of improved system's reachable states**

What is needed now is a methodology for converting behavioral rules defined by CSM models into object-oriented program constructs. This issue exceeds the scope of the present paper, we should note only that the research on it is now under progress and we expect that a new type of the tool supporting the analysis and design of concurrent software using CSM methodology will be operable soon.